\documentclass[%
reprint,
superscriptaddress,
frontmatterverbose,
amsmath,amssymb,
aps,
prx
longbibliography 
]{revtex4-2}

\usepackage{graphicx}
\usepackage{dcolumn}
\usepackage{bm}
\usepackage[mathlines]{lineno}
\usepackage{braket}
\usepackage{color}
\usepackage{times}

\begin{document}

\preprint{NC/123-Singularities}

\title{Cubic singularities in binary linear electromechanical oscillators}

\author{Xin Zhou}
\affiliation{College of Intelligence Science and Technology, NUDT, Changsha 410073, China}%
\affiliation{These authors contributed equally: Xin Zhou, Hui Jing, Xingjing Ren.}

\author{Hui Jing}
\affiliation{Key Laboratory of Low-Dimensional Quantum Structures and Quantum Control of Ministry of Education, Department of Physics and Synergetic Innovation Center for Quantum Effects and Applications, Hunan Normal University, Changsha 410081, China}%
\affiliation{These authors contributed equally: Xin Zhou, Hui Jing, Xingjing Ren.}

\author{Xingjing Ren}
\affiliation{College of Intelligence Science and Technology, NUDT, Changsha 410073, China}%
\affiliation{These authors contributed equally: Xin Zhou, Hui Jing, Xingjing Ren.}

\author{Jianqi Zhang}
\affiliation{State Key Laboratory of Magnetic Resonance and Atomic and Molecular Physics,
Wuhan Institute of Physics and Mathematics, Innovation Academy of Precision Measurement Science and Technology, Chinese Academy of Sciences, Wuhan 430071, China}%

\author{Ran Huang}
\affiliation{Theoretical Quantum Physics Laboratory, Cluster for Pioneering Research, RIKEN, Wako-shi, Saitama 351-0198, Japan}%

\author{Zhipeng Li}
\affiliation{Department of Electrical and Computer Engineering, National University of Singapore, Singapore 117576, Singapore.}%

\author{Xiaopeng Sun}
\affiliation{College of Intelligence Science and Technology, NUDT, Changsha 410073, China}%

\author{Xuezhong Wu}
\affiliation{College of Intelligence Science and Technology, NUDT, Changsha 410073, China}%

\author{Cheng-Wei Qiu}
\affiliation{Department of Electrical and Computer Engineering, National University of Singapore, Singapore 117576, Singapore.}%

\author{Franco Nori}%
\affiliation{Theoretical Quantum Physics Laboratory, Cluster for Pioneering Research, RIKEN, Wako-shi, Saitama 351-0198, Japan}%

\author{Dingbang Xiao}%
\affiliation{College of Intelligence Science and Technology, NUDT, Changsha 410073, China}%

\date{\today}

\begin{abstract}
Singularities arise in diverse disciplines and play a key role in both exploring fundamental laws of physics and making highly-sensitive sensors \cite{Arnold1984Catastrophe,Liu2016Metrology,Chen2017Exceptional,Lai2019Observation,Hokmabadi2019Non-Hermitian, Kononchuk2022Exceptional, Mohammad2019Exceptional,Ozdemir2019Parity-time,Doppler2016Dynamically,Xu2016Topological,Hassan2017Dynamically,Yoon2018Time,Zhang2018Dynamically, Nasari2022Observation,Mohammad2019Exceptional, Ozdemir2019Parity-time}. Higher-order ($\ge 3$) singularities, with further improved performance \cite{Hodaei2017Enhanced,Tang2020Exceptional,delPino2022NonHermitian,Wang2019Dynamics,Bai2022Nonlinearity}, however, usually require exquisite tuning of multiple ($\ge 3$) coupled degrees of freedom \cite{Hodaei2017Enhanced,Tang2020Exceptional,delPino2022NonHermitian} or nonlinear control \cite{Wang2019Dynamics,Bai2022Nonlinearity}, thus severely limiting their applications in practice. Here we propose theoretically and confirm using mechanics experiments that, cubic singularities can be realized in a coupled binary system without any nonlinearity, only by observing the phase tomography of the driven response. By steering the cubic phase-tomographic singularities in an electrostatically-tunable micromechanical system, enhanced cubic-root response to frequency perturbation and voltage-controlled nonreciprocity are demonstrated. Our work opens up a new phase-tomographic method for interacted-system research, and sheds new light on building and engineering advanced singular devices with simple and well-controllable elements, with a wide range of applications including precision metrology, portable nonreciprocal devices, and on-chip mechanical computing.
\end{abstract}

\maketitle


Singularities, sometimes referred to as catastrophes, arise in diverse disciplines and play an essential role in describing how the properties of an object, that are dependent on certain controlling parameters, change qualitatively even if the controlling parameters vary minimally \cite{Arnold1984Catastrophe}. The unusual landscapes near the singularities are very useful for enhancing the sensitivities of detection \cite{Liu2016Metrology,Chen2017Exceptional,Lai2019Observation,Hokmabadi2019Non-Hermitian, Kononchuk2022Exceptional,Mohammad2019Exceptional,Ozdemir2019Parity-time} as well as generating nonreciprocity \cite{Doppler2016Dynamically,Xu2016Topological,Hassan2017Dynamically, Yoon2018Time,Zhang2018Dynamically,Nasari2022Observation,Mohammad2019Exceptional, Ozdemir2019Parity-time}. Recently, higher-order singularities have increasingly attracted attention, which have the potential to provide higher performance and engender richer physics \cite{Hodaei2017Enhanced,Tang2020Exceptional,delPino2022NonHermitian, Wang2019Dynamics,Bai2022Nonlinearity}. However, what prevents the higher-order singularities from being well explored or exploited thus far is their difficulty in practical realization and control. Usually, higher-order singularities call for multiple ($\ge 3$) degrees of freedom \cite{Hodaei2017Enhanced,Tang2020Exceptional,delPino2022NonHermitian} and are very difficult to construct and adjust. Lately, it was theoretically predicted \cite{Wang2019Dynamics} and experimentally observed \cite{Bai2022Nonlinearity} that introducing nonlinearity to binary non-Hermitian systems may also realize higher-order singularities. These studies point out the possibility of exploring higher-dimensional ($\ge 3$) physics using binary systems. However, nonlinearity is still a stringent condition that will bring intrinsic power consumption and reliability limits.

Micro- and nanomechanical resonators, with broad applications \cite{Ng2015The,Nguyen1998Micromachined,Hanay2012Single,Cleland1998Ananometre-scale, Middlemiss2016Measurement,Ayazi2001HARPSS}, excellent in-situ controllability \cite{Faust2012Nonadiabatic,Mahboob2012Phonon,Zhou2019NC, Eriksson2023Controllable}, and rich interactive phenomena \cite{Faust2012Nonadiabatic,Mahboob2012Phonon,Okamoto2013Coherent,Faust2013Coherent, Sun2016Correlated,Zhou2019NC,Miao2022Nonlinearity,Xu2016Topological}, provide an ideal platform for exploring and exploiting singularities. Here we demonstrate, both theoretically and experimentally, that cubic singularity arcs and nexus can be realized in a binary linear micromechanical system, by observing the phase-locked-loop (PLL) enabled tomographic dynamics of the coherent-coupling phase responses. The cubic singularity arcs are featured by a series of stable boundaries on a partially folded geometry made by the closed-loop oscillation frequency of a pair of coherently coupled modes. The intersection of two singularity arcs makes a singularity nexus. Behind the singularities lies interesting polarization dynamics. In an electrically tunable micromechanical system, we experimentally observe the cubic singularities. We confirm that the singularity nexus can provide improved detecting sensitivity with a cubic-root response, surpassing the binary singularities. Moreover, we demonstrate a single-parameter-controlled nonreciprocity by traversing the projected sensitivity arcs. Using the electrostatic tuning method, the nonreciprocity is steered electrically.

\begin{figure*}
\centering
\includegraphics[width=180mm]{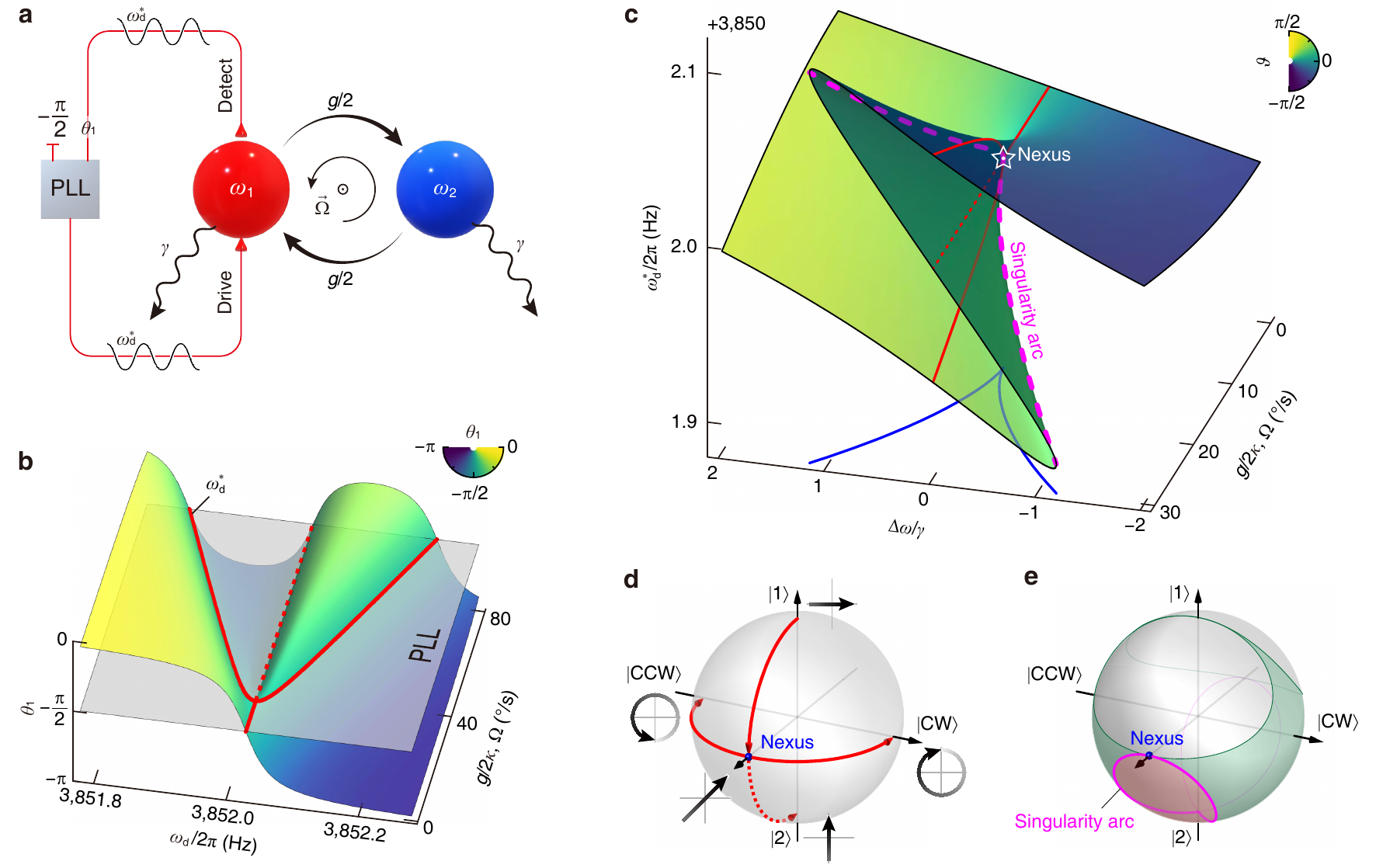}
\caption{\label{fig:Sing} \textbf{Cubic singularity in phase-tomographic dynamics: Concept.}
\textbf{a} Setup for the phase-tomographic singularity. Mode~1 is driven and coherently coupled to mode~2. A PLL is used to obtain a phase-tomographic closed-loop oscillation. In this study, coherent coupling is produced by rotation $\vec{\Omega}$, giving a coupling strength of $g = 2 \kappa \Omega$ with $\kappa=0.85$.
\textbf{b} Open-loop phase-frequency response ($\theta_1$, colored surface) of the driven mode~1 as a function of coupling strength. PLL adjusts the drive frequency to track the phase $\theta_1 = - \pi/2$. The phase tomography shows a ``pitchfork'' bifurcation.
\textbf{c} The $\omega_\text{d}^*$ as a function of degeneracy condition $\Delta \omega$ and coupling strength $g$. Singularity arcs connected by a singularity nexus are formed by the stability boundaries. Colors on the surface represent the relative phase of the coupled modes.
\textbf{d} Polarization dynamics of the balanced ``pitchfork'' bifurcation on the classical Bloch sphere.
\textbf{e} Stabilities and singularities on the Bloch sphere.
}
\end{figure*}

\begin{figure*}
\centering
\includegraphics[width=180mm]{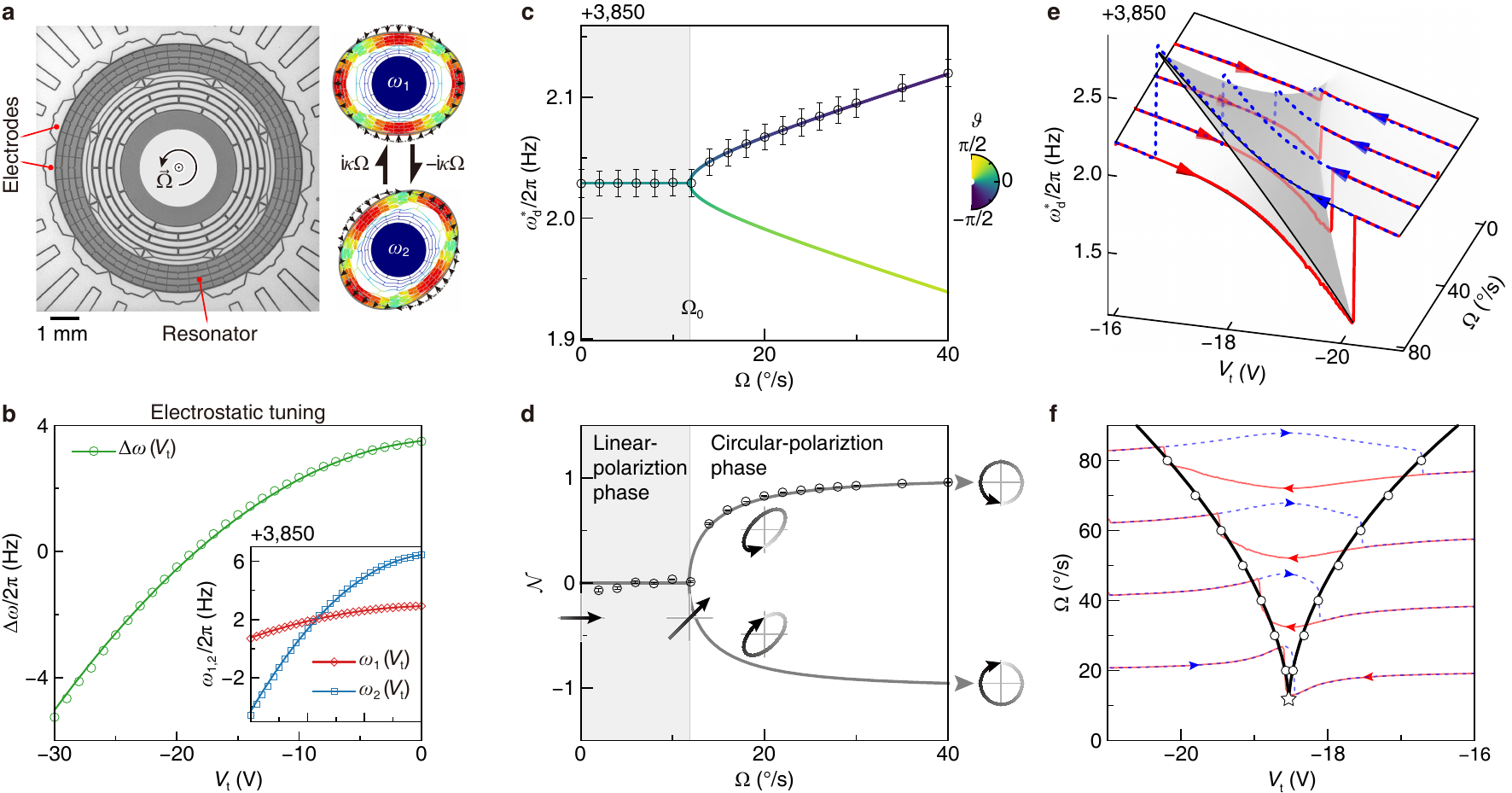}
\caption{\label{fig:Surf} \textbf{Experimental realization of the phase-tomographic singularity.}
\textbf{a} Micromechanical resonator with degenerate modes that are coherently coupled by Coriolis effect.
\textbf{b} Mode natural frequencies $\omega_{1,2}$ and their difference $\Delta \omega$ vs tuning voltage $V_\text{t}$.
\textbf{c} Spontaneous chiral symmetry breaking and \textbf{d} the corresponding order parameter $\mathcal{N}$ illustrating a second-order phase transition. Error bars are the standard deviation.
\textbf{e} Measured $\omega_\text{d}^*$ if the tuning voltage $V_\text{t}$ is swept adiabatically at constant angular velocities. The blue dashed (red solid) curves indicate the upward (downward) sweeps. Singularities and hysteresis are shown above $\Omega_0$. The gray surface is simulation.
\textbf{f} Cubic singularities projected to $V_\text{t}$-$\Omega$ plane. White-faced points (black curves) are experimental (theoretical) data. The red and blue curves are the corresponding $\mathcal{N}$ data in \textbf{e} revealing first-order phase transitions.
}
\end{figure*}

\section*{Results}
\paragraph*{\textbf{Concept:}}

We consider a pair of coherently coupled mechanics modes with tunable natural frequencies $\omega_{1,2}$ and identical dissipation rate $\gamma$, as shown in Fig.~\ref{fig:Sing}a. In this study, the coherent coupling is produced by the rotation-induced Coriolis effect \cite{Li2018Coriolis} (see Supplementary note 2), giving an angular velocity $\Omega$-controlled coupling strength $g =2 \kappa \Omega$, where $\kappa \approx 0.85$ is the Coriolis coupling coefficient. Mode~1 is linearly driven by an external sinusoidal force with frequency $\omega_\text{d}$ while mode~2 is set free. The linear displacement responses of the two modes are labelled by $q_{1,2}= \cos(\omega_\text{d}t + \theta_{1,2})$, where $\vert q_{1,2} \vert$ ($\theta_{1,2}$) are the amplitude (phase) responses.

If the two modes are at degeneracy ($\Delta \omega \equiv \omega_2 -\omega_1 =0$), the open-loop amplitude-frequency response $\vert q_1 \vert$ of mode~1 as a function of the coupling strength $g$ displays normal-mode splitting \cite{Faust2012Nonadiabatic} (see Supplementary Fig.~4 or 5). The accompanied phase response $\theta_1$ of mode~1 is shown by the colored surface in Fig.~\ref{fig:Sing}b. Here, we consider the tomography of the driven-mode phase response with a constant oscillation phase, which is $-\pi/2$ for ideal oscillators. The phase tomography is realized by imposing a PLL, which produces a stable closed-loop oscillation (Fig.~\ref{fig:Sing}a). As shown by the red contour in Fig.~\ref{fig:Sing}b, the controlled closed-loop frequency (denoted by $\omega_\text{d}^*$, that fulfills the phase-tomographic condition $\theta_1 = -\pi/2$) as a function of coupling strength shows a ``pitchfork'' bifurcation. It is noteworthy that this bifurcation is only related to the landscape of the linear phase response $\theta_1$, and is different from its counterparts in nonlinear dynamics \cite{Strogatz2000Nonlinear}. The degenerate bifurcation point is exactly the threshold between weak and strong coupling, $g=\gamma$. The middle branch of the bifurcation (dotted curves) is unstable because the corresponding amplitude response is in an antiresonance valley (see Supplementary Fig.~4 or 5), which is unlikely to be detected by the PLL. The other two branches (solid curves) are stable because the corresponding amplitude responses are close to the resonant peaks.

The ``pitchfork'' bifurcation of $\omega_\text{d}^*$ changes as the degeneracy condition $\Delta \omega$ varies. As a function of the coupling strength $g$ and degeneracy condition $\Delta \omega$, $\omega_\text{d}^*$ makes a 3D surface (Fig.~\ref{fig:Sing}c), which is simulated by the cubic equation (see Supplementary note 3)
\begin{align}\label{equ:PLock}
(\omega_\text{d}^*-\omega_1) (\omega_\text{d}^*-\omega_2+\frac{\text{i}}{2}\gamma)(\omega_\text{d}^*-\omega_2-\frac{\text{i}}{2}\gamma) - \frac{1}{4}g^2 (\omega_\text{d}^*-\omega_2) = 0.
\end{align}
The inflectional part of the folded $\omega_\text{d}^*$ surface is unstable. If the stability boundaries (magenta curves) are crossed by changing the control parameters $g$ and $\Delta \omega$ in an adiabatic manner, catastrophic jumps of the oscillation state takes place. The stability boundaries constitute singularity arcs. The degenerate bifurcation point (white star) is actually a singularity nexus that connects two singularity arcs, giving a severely twisted $\omega_\text{d}^*$ geometry. The singularity arcs and nexus, given the discriminant of the cubic Eq.~\ref{equ:PLock}, construct cubic cusp catastrophes, the projection of which to the $\Delta \omega$-$g$ parameter plane forms two cusp-connected parabola loci \cite{Arnold1984Catastrophe,Strogatz2000Nonlinear} (see Supplementary note 4).

The phase-tomographic closed-loop oscillation of the system is described by a state vector $\vert \psi \rangle = \cos \frac{\phi}{2} \vert 1 \rangle +\text{e}^{i \vartheta} \sin \frac{\phi}{2}  \vert 2 \rangle$, where $\{\vert 1 \rangle,\vert 2 \rangle \}$ is the orthonormal basis of modes 1 and 2, $\phi = 2 \arctan (\lvert q_2 \rvert/\lvert q_1 \rvert)$ is the polar angle, and $\vartheta \equiv \theta_2 - \theta_1$ is the relative phase between the coupled modes, which is revealed by the colors of the surface in Fig.~\ref{fig:Sing}c. The state vectors can be projected to a classical Bloch sphere with polar and azimuthal angles given by $\phi$ and $\vartheta$, respectively (see Supplementary note 5). The red trajectories on the Bloch sphere in Fig.~\ref{fig:Sing}d illustrate the dynamics that underlies the degenerate ``pitchfork'' bifurcation in Fig.~\ref{fig:Sing}b. The arrows indicate the $g$-increasing direction. There is a corresponding polarization pattern in the $q_1$-$q_2$ plane for every state vector on the Bloch sphere. At the initial state $\vert 1 \rangle$ without coupling, $q_2=0$ makes a horizontal linear polarization. When $\Omega \to \infty$, the states corresponding to the upper or lower stable branches of $\omega_\text{d}^*$ approach the counter-clockwise circular polarization state $\vert \text{CCW} \rangle = (\vert 1 \rangle- i \vert 2 \rangle)/\sqrt{2}$, or the clockwise circular polarization state $\vert \text{CW} \rangle = (\vert 1 \rangle + i \vert 2 \rangle)/\sqrt{2}$, respectively. While the state corresponding to the middle unstable branch approaches $\vert 2 \rangle$ with vertical linear polarization, because the antiresonance valley makes $q_1 \to 0$. At the singularity nexus, $\vert q_1 \vert = \vert q_2 \vert$ and $\vartheta = 0$ make a $45^\circ$ linear polarization. The singularity arcs are marked by the magenta curves on the Bloch sphere in Fig.~\ref{fig:Sing}e. The unstable, bistable, and monostable regions on the Bloch sphere are filled with light red, light green, and grey, respectively. The front (back) hemisphere corresponds to the oscillations with positive (negative) angular velocities. The binary polarization dynamics is a projection of a cubic singular dynamics.

\begin{figure*}
\centering
\includegraphics[width=120mm]{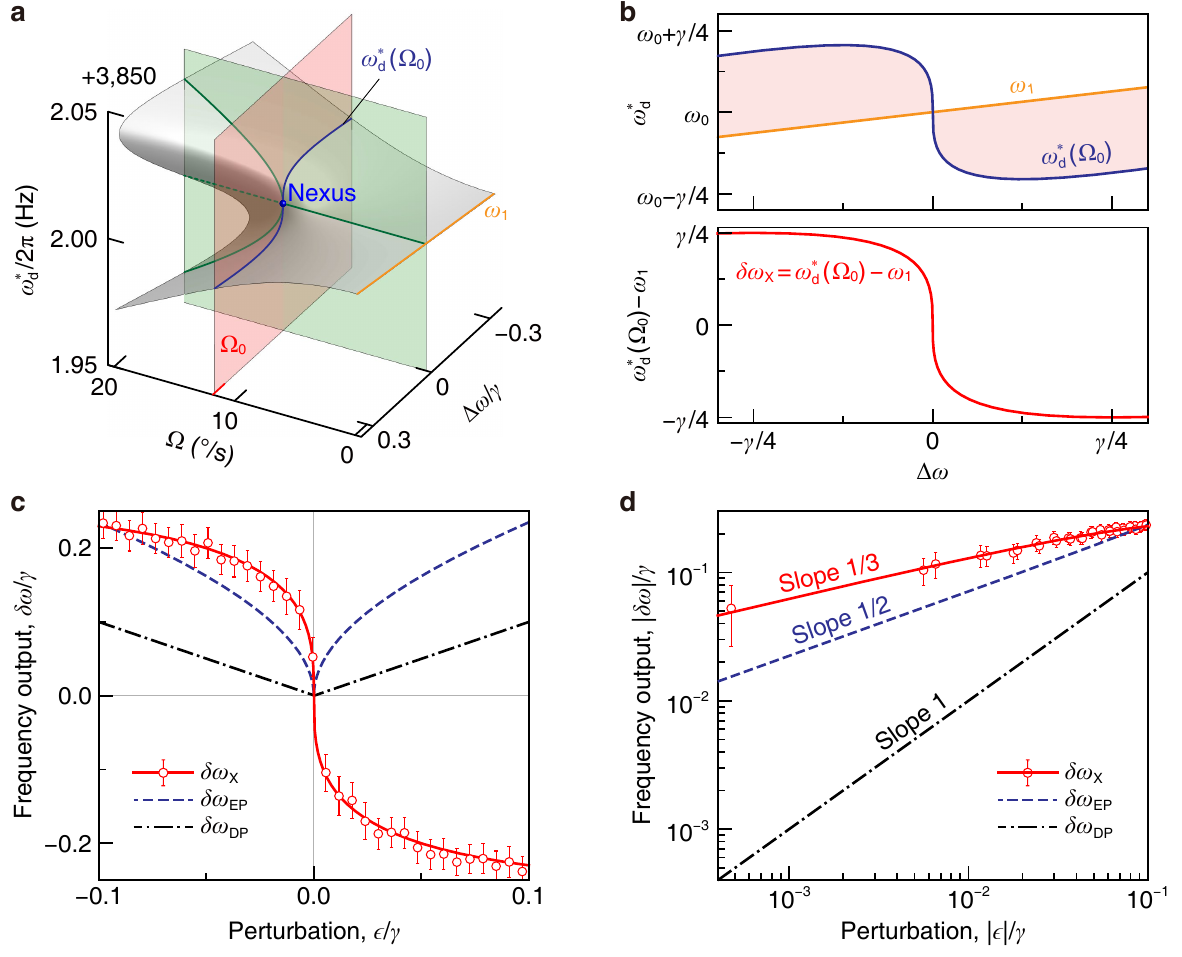}
\caption{\label{fig:Pro} \textbf{High sensitivity near the singularity nexus.}
\textbf{a} Phase-tomographic frequency $\omega_\text{d}^*$ as a function of angular velocity $\Omega$ and degeneracy condition $\Delta \omega$. The contours of $\Omega = \Omega_0$ (blue curve) and $\Delta \omega = 0$ (green curves) portray the sharp variation of $\omega_\text{d}^*$ at the singularity nexus.
\textbf{b} Phase-tomographic frequency at the nexus angular velocity $\omega_\text{d}^* (\Omega_0)$ and its shift from $\omega_1$, $\delta \omega_\text{X} = \omega_\text{d}^* (\Omega_0) -\omega_1$ as functions of $\Delta \omega$. $\omega_0$ denotes $\omega_1$ at $\Delta \omega = 0$. In the range of $-0.25 \gamma \le \Delta \omega \le 0.25 \gamma$, $\delta \omega_\text{X}$ decreases monotonically to $\Delta \omega$.
\textbf{c} Frequency output $\delta \omega_\text{X}$ near the singularity nexus versus the natural-frequency perturbation $\epsilon = \Delta \omega$ from simulation (red solid curve) and experiment (points). Eigenfrequency splits near an EP (blue dashed curve) and a DP (black dot-dashed curve) are simulated as well. Error bars are the standard deviation.
\textbf{d} Logarithmic plot of the absolute data in \textbf{c}. The singularity nexus has a cubic-root output, which provides higher sensitivity than the EP and DP.
}
\end{figure*}

\begin{figure*}
\centering
\includegraphics[width=120mm]{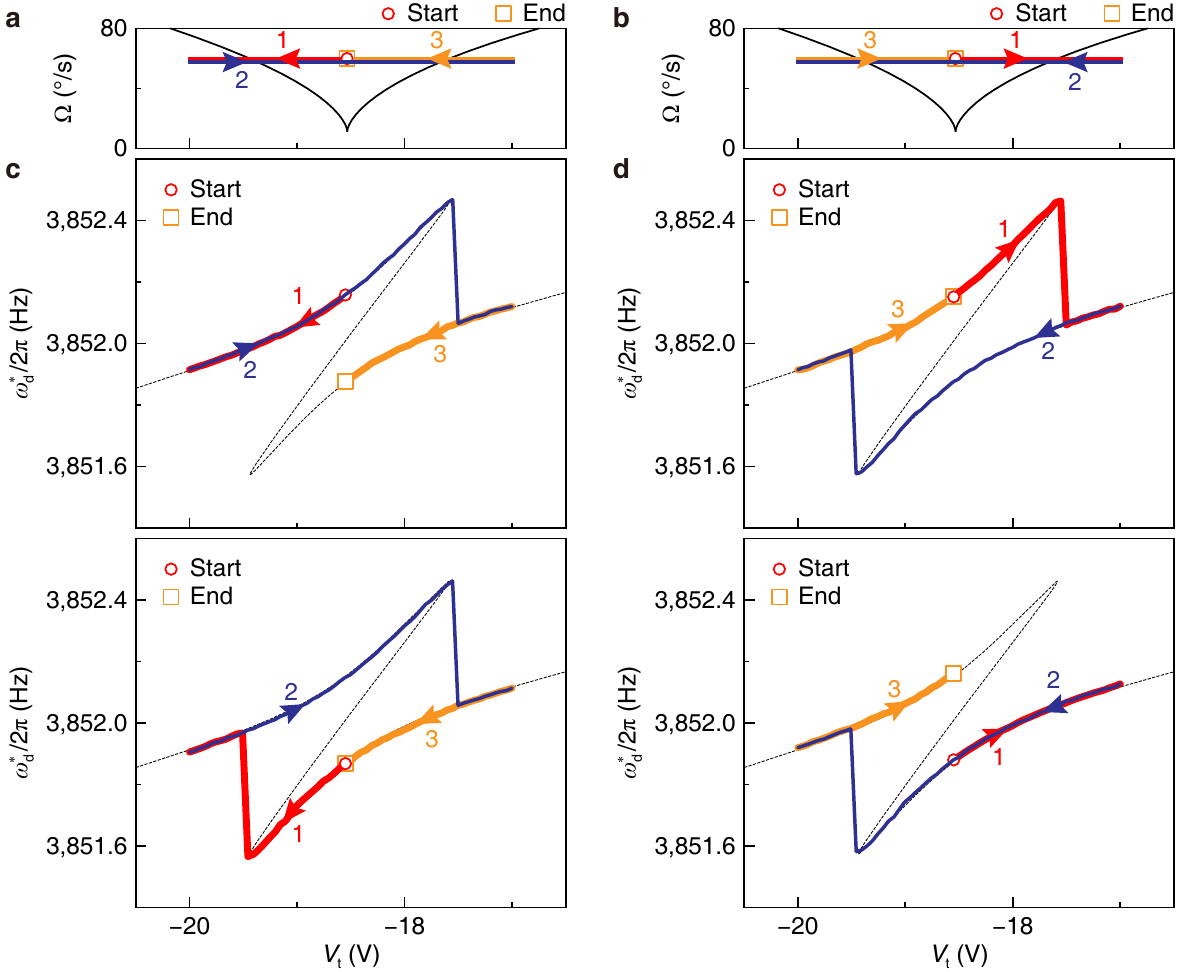}
\caption{\label{fig:NR} \textbf{Voltage controlled nonreciprocity.}
A closed 1D parameter trajectory that crosses both projected singularity loci in down-up-down \textbf{a} and up-down-up \textbf{b} $V_\text{t}$ traversing process. The numbers indicate the traversing order.
\textbf{c} Adiabatic evolutions start at high (upper panel) or low (lower panel) bistable branches following the down-up-down $V_\text{t}$ traversing process.
\textbf{d} Same as \textbf{c} for the up-down-up traversing process. Two processes that start at an identical location and travel the same path in opposite directions will reach distinct destinations. The nonreciprocity is ensured by crossing both sides of the projected singularity loci.
}
\end{figure*}

\paragraph*{\textbf{Realization:}}
To experimentally demonstrate the cubic singularities, we realize the scheme in Fig.~\ref{fig:Sing}a using a capacitive microelectromechanical system \cite{Zhou2019NC} (see Supplementary note 1). As shown in Fig.~\ref{fig:Surf}a, two micromechanical modes with near-degenerate natural frequencies $\omega_{1,2}/2\pi \approx 3.85$~kHz and equal dissipation rates $\gamma = 2\pi \times 55.8$~mHz are coherently coupled by the Coriolis effect. The degeneracy condition $\Delta \omega$ can be adjusted electrostatically by a tuning voltage $V_\text{t}$, as shown in Fig.~\ref{fig:Surf}b (see Methods). Mode~1 is driven externally and the linear displacement responses of the two modes are detected using the Homodyne method (see Methods). A stable $\theta_1= -\pi/2$ phase-tomographic closed-loop oscillation is maintained by applying a PLL to mode~1.

First, we adjust $V_\text{t}$ to make $\Delta \omega \approx 0$ and adiabatically sweep $\Omega$ from zero to $40^\circ/\text{s}$ to observe the singularity nexus. The calculated and measured phase-tomographic closed-loop frequency $\omega_\text{d}^*$ is shown in Fig.~\ref{fig:Surf}c. If the angular velocity is below the strong-coupling threshold $\Omega < \gamma/(2 \kappa)$, $\omega_\text{d}^*$ is latched to $\omega_1$. At the threshold (singularity nexus), $\Omega_0 = \gamma/(2 \kappa) = 11.83^\circ/\text{s}$, the $\omega_\text{d}^*$ transfers to one of the two stable bifurcation branches randomly, revealing a spontaneous breaking of chiral symmetry. To better describe the chirality, we re-expand the state vector in the $\{\vert \text{CCW} \rangle,\vert \text{CW} \rangle \}$ basis, $\vert \psi \rangle = c_\text{ccw} \vert \text{CCW} \rangle +c_\text{cw}  \vert \text{CW} \rangle$, and define the relative population of the $\vert \text{CCW} \rangle$ and $\vert \text{CW} \rangle$ states as the order parameter $\mathcal{N} = (\vert c_\text{ccw} \vert^2 - \vert c_\text{cw} \vert^2)/(\vert c_\text{ccw} \vert^2 + \vert c_\text{cw} \vert^2) = - \sin \phi \sin \vartheta$ (see Supplementary note 6). The order-parameter evolution in the symmetry-breaking process of Fig.~\ref{fig:Surf}c is plotted in Fig.~\ref{fig:Surf}d, showing a second-order phase transition. Below the singularity nexus, the oscillation is in the linear-polarization phase. At the nexus, the oscillation transfers to one of the degenerate $\vert \text{CCW} \rangle$ and $\vert \text{CW} \rangle$ dominant phases randomly. The oscillation frequency of the $\vert \text{CCW} \rangle$ ($\vert \text{CW} \rangle$) dominant phase increases (decreases) if $\Omega$ is further increased from $\Omega_0$, because of the rotational Doppler effect \cite{Romain2014Sound,Deng2021Rotational}. The experimental data in Fig.~\ref{fig:Surf}c and d indicate that we experimentally observed a linear-polarization-to-$\vert \text{CCW} \rangle$ phase transition.

Next, we change $\Delta \omega$ by tailoring $V_\text{t}$ adiabatically while keeping $\Omega$ unchanged at some specific values to observe the singularity arcs. The measured $\omega_\text{d}^*$ is shown in Fig.~\ref{fig:Surf}e. The blue dashed (red solid) curves illustrate the experimental results for upward (downward) $V_\text{t}$ sweeps. If $\Omega > \Omega_0$, the sweeping curves encounter discontinuous jumps at some specific values of $V_\text{t}$, referred to as catastrophes or singularities \cite{Arnold1984Catastrophe}. The two upward and downward curves at identical $\Omega$ form a hysteresis loop. The area of the hysteresis loop decreases when reducing $\Omega$, which vanishes if $\Omega \le \Omega_0$. The experimentally detected singularities mapped to the $V_\text{t}$-$\Omega$ parameter plane develop the predicted cusp-connected parabola loci in Fig.~\ref{fig:Sing}c, as shown in Fig.~\ref{fig:Surf}f (see Supplementary note 4). The order parameters $\mathcal{N}$ corresponding to the upward (downward) sweeps in Fig.~\ref{fig:Surf}e are shown by the blue dashed (red solid) curves in Fig.~\ref{fig:Surf}f, which indicate first-order transitions from the $\vert \text{CCW} \rangle$ ($\vert \text{CW} \rangle$) dominant phase to the $\vert \text{CW} \rangle$ ($\vert \text{CCW} \rangle$) dominant phase.

\paragraph*{\textbf{Cubic-root sensitivity:}}
We now demonstrate the enhanced cubic-root sensitivity of the singularity nexus. As shown in Fig.~\ref{fig:Pro}a, at the nexus ($\Omega = \Omega_0$ and $\Delta \omega = 0$), the closed-loop oscillation frequency $\omega_\text{d}^* (\Omega_0)$ is latched to the driven-mode natural frequency $\omega_1$. Otherwise, if the degeneracy is broken, $\Delta \omega \ne 0$, $\omega_\text{d}^* (\Omega_0)$ will suddenly but continuously deviate from $\omega_1$. The deviation $\delta \omega_\text{X} = \omega_\text{d}^* (\Omega_0) - \omega_1$ changes sharply if $\Delta \omega$ shift from the nexus (Fig.~\ref{fig:Pro}b). Here, we consider the perturbation $\epsilon$ that can affect the degeneracy condition, $\epsilon \sim \Delta \omega$, and regard $\delta \omega_\text{X}$ in the vicinity of the nexus as the sensing output of $\epsilon$, as shown by the red curve in Fig.~\ref{fig:Pro}c (see Supplementary note 7). When plotted on a logarithmic scale, it gives a cubic-root response near the nexus: $\delta \omega_\text{X} \sim \epsilon^{1/3}$ (Fig.~\ref{fig:Pro}d), confirming the cubic nature of the singularity nexus. To experimentally verify the cubic-root behavior, we maintain an $\Omega_0$ rotation and introduce a fine-tuning voltage $V_\text{t}$ to sweep across the nexus. By transforming $V_\text{t}$ to $\epsilon$, the experimental input-output data are shown by the red circles in Fig.~\ref{fig:Pro}c and d, which coincide well with the cubic-root simulation.

We compare the sensitivities produced by the singularity nexus and a binary exceptional-point (EP) singularity that is generated by a passive parity-time-symmetric system whose damping difference is chosen to be equal to the dissipation of our system (see Supplementary note 7). The blue dashed curves in Fig.~\ref{fig:Pro}c and d demonstrate an $\epsilon^{1/2}$ dependency of the eigenfrequency split $\delta \omega_\text{EP}$ near the EP. The sensitivity produced by the singularity nexus is greater than that of the binary EP \cite{Liu2016Metrology,Chen2017Exceptional,Lai2019Observation,Hokmabadi2019Non-Hermitian} and is on par with that of the third-order EP \cite{Hodaei2017Enhanced}. When compared to the standard output $\delta \omega_\text{DP} \sim \epsilon$ of the diabolic-point (DP) system shown by the black dot-dashed curves in Fig.~\ref{fig:Pro}c and d, both $\delta \omega_\text{EP}$ and $\delta \omega_\text{X}$ are much improved.

\paragraph*{\textbf{Voltage-controlled nonreciprocity:}}
Lastly, we show that the phase-tomographic cubic singularity can produce a voltage controlled nonreciprocity. We consider a closed 1D trajectory along the tuning-voltage $V_\text{t}$ direction in the parameter plane, which starts and ends at the bistable region and crosses both projected singularity loci. If the trajectory is traversed in the down-up-down (up-down-up) direction, as shown in Fig.~\ref{fig:NR}a (b), the system ends at the low (high) branch no matter which branch it is started, as shown in Fig.~\ref{fig:NR}c (d). Even if it is started at an identical location, opposite traversal directions lead to different ending branches, as shown in Fig.~\ref{fig:NR}c or d. The ending location only depends on the traversal direction, not the starting place.

In the experiments shown in Fig.~\ref{fig:NR}, the tuning voltage $V_\text{t}$ is the only steering knob of the nonreciprocal state transfer, while the angular velocity is set to be a constant value, $\Omega = 60^\circ/\text{s}$. In fact, the $\Omega$ value of the $V_\text{t}$-controlled nonreciprocal process can be changed almost arbitrarily, as long as the start/end points are located at the bistable region and both projected singularity loci are crossed. This single-parameter-controlled nonreciprocity is more desirable than that of the binary EP singularity, which is guaranteed by two-parameter encircling \cite{Gao2015Observation,Doppler2016Dynamically,Xu2016Topological, Hassan2017Dynamically,Yoon2018Time,Zhang2018Dynamically,Nasari2022Observation}. Benefiting from the electrostatic tunability of our device, the phase-tomographic cubic singularity provides a voltage-controlled nonreciprocity.

\section*{Discussion}
Although we experimentally demonstrate the phase-tomographic closed-loop cubic singularity based on the Coriolis coupling, in principle, it can also be realized using ordinary linear coherent coupling (see Supplementary discussion). This study may open up a new tomographic dimension for phase-related interactive dynamics study, which is applicable for a wide range of disciplines such as optics, optomechanics, or hybrid quantum systems. Our discovery makes it possible to construct advanced singularities using highly controllable elements. It also enhances the understanding of the closed-loop oscillation dynamics, and extends coherent control into the singularity region. Potential applications of the closed-loop singularity include precise sensing, deep-sub-linewidth mode matching, rapid mode switching, and generating portable nonreciprocity. Moreover, the hysteresis in the closed-loop oscillation is promising for mechanical computing \cite{Yasuda2021Mechanical}. The bit abstraction of the closed-loop oscillations is independent of vibration amplitudes, which may provide potential advantages in power consumption and lifetime. Future studies can also investigate phase-tomographic singularities originating from different kinds of coupling, the interplay with other kinds of singularities, and the phase-tomographic dynamics in many-body systems with more degrees of freedom \cite{Matheny2019Exotic,delPino2022NonHermitian,Charles2022Direct}.

\section*{\label{sec:Method}Methods}

\paragraph*{\textbf{Electrostatic frequency tuning:}}

The tuning voltage $V_\text{t}$ introduces electrostatic negative stiffness to both modes~1 and 2, $\omega^2_{1,2} (V_\text{t})= {\omega'}_{1,2}^{2}(0) - T_{1,2} (V_0 - V_\text{t})^2$, where $\omega'_{1,2} (0)$ denotes the natural frequencies of the bare mechanical modes. The electrostatic tuning factors $T_{1,2}$ are proportional to the capacitive area, the inverse of the modal mass, and the inverse of the cubic of the capacitive gap. The stiffness perturbation induced by $V_0$ exists even if the tuning voltage $V_\text{t}$ is absent, so it can be included in the intrinsic natural frequencies. By defining $\omega_{1,2}^2 (0) = {\omega'}^{2}_{1,2}(0) - T_{1,2}V_0^2$, and assuming that the electrostatic stiffness perturbation is small relative to the intrinsic stiffness, we have
\begin{align}
\omega_{1,2} (V_\text{t}) &=\sqrt{ \omega_{1,2}^2 (0) + T_{1,2} (2 V_0 V_\text{t} -V_\text{t}^2)} \notag \\
&\approx \omega_{1,2} (0) + K_{1,2}  (2 V_0 V_\text{t} -V_\text{t}^2), \label{equ:ft0}
\end{align}
where the tuning coefficients are defined by $K_{1,2} = T_{1,2}/[2 \omega_{1,2}(0)]$.

The experimentally measured natural frequencies $\omega_{1,2}$ as functions of $V_\text{t}$ are shown by the red and blue points in the inset of Fig.~\ref{fig:Surf}b, which are fitted (curves) to the model (\ref{equ:ft0}) with parameters $\omega_1 (0) = 2\pi \times 3,852.92$~Hz, $\omega_2 (0) = 2\pi \times 3,856.43$~Hz, and $V_0=2.5~\text{V}$. The tuning coefficients are fitted to be $K_1 = 1.29 \times 10^{-2}~\text{rad}~\text{s}^{-1}~\text{V}^{-2}$ and $K_2 = 6.40\times 10^{-2}~\text{rad}~\text{s}^{-1}~\text{V}^{-2}$. The relationship between the difference of natural frequencies (degeneracy condition) $\Delta \omega = \omega_2-\omega_1$ and the tuning voltage $V_\text{t}$ is further given by
\begin{align}\label{equ:ft1}
\Delta \omega (V_\text{t}) &= \Delta \omega (0) + ( K_2 - K_1)(2 V_0 V_\text{t} -V_\text{t}^2),
\end{align}
where $\Delta \omega (0) = \omega_2 (0)- \omega_1 (0)$. The experimentally measured $\Delta \omega$ data (green points in Fig.~\ref{fig:Surf}b) coincide well with the tuning model (\ref{equ:ft1}) (green curve in Fig.~\ref{fig:Surf}b).

\paragraph*{\textbf{Homodyne measurement:}}

The antinodal displacements of the two micromechanical modes $q_{1,2} = \lvert q_{1,2} \rvert \cos(\omega_\text{d}t+ \theta_{1,2})$ are picked up by capacitive transducers, transformed into voltage signals by two charge amplifiers integrated into a printed circuit board, and then recorded by a two-channel lock-in amplifier (Zurich Instruments HF2LI). The amplitudes $\lvert q_{1,2} \rvert$ and phases $\theta_{1,2}$ relative to the driving signal are obtained by dual-phase demodulation techniques. In this process, $q_j(\omega_\text{d},t)$ is split and separately mixed with the driving reference signal $\cos{\omega_\text{d}t}$ and a $\pi/2$-phase-shifted copy of it,
\begin{align*}
&\lvert q_j \rvert \cos(\omega_\text{d}t+ \theta_j)  \times  \cos{\omega_\text{d}t} \\ \notag
= &\frac{\lvert q_j \rvert }{2}\left[ \cos(2 \omega_\text{d}t+ \theta_j) + \cos(\theta_j) \right], \\
&\lvert q_j \rvert \cos(\omega_\text{d}t+ \theta_j)  \times  \cos{(\omega_\text{d}t + \frac{\pi}{2})}  \\ \notag
=& \frac{\lvert q_j \rvert }{2}\left[ -\sin(2 \omega_\text{d}t+ \theta_j) + \sin(\theta_j) \right].
\end{align*}
The high-harmonic components of the mixed signals are removed using low-pass filters and the remaining in-phase component $X_j= \frac{\lvert q_j \rvert }{2} \cos(\theta_j) $ and quadrature component $Y_j= \frac{\lvert q_j \rvert }{2} \sin(\theta_j)$ are obtained. By transforming into the polar coordinates, we can obtain the amplitude and phase,
\begin{align*}
\lvert q_j \rvert &= \sqrt{X^2+Y^2}, \\
\theta_j &= \arctan \frac{Y}{X}.
\end{align*}

\section*{\label{sec:Data}Data availability}
Data relevant to the figures and conclusions of this manuscript are available at https://doi.org/10.6084/m9.figshare.19609350.


%

\begin{acknowledgments}
X.Z. thank Prof. Ashwin Seshia from University of Cambridge and Prof. Zenghui Wang from University of Electronic Science and Technology of China for helpful discussions. This work is partly supported by the National Natural Science Foundation of China (NSFC) (U21A20505 and 51905539), the Young Elite Scientist Sponsorship Program by CAST (YESS20200127), and the Natural Science Foundation of Hunan Province for Excellent Young Scientists (2021JJ20049). This work is primarily supported the National Key R\&D Program of China (NKPs) (2022YFB3204901).

\end{acknowledgments}

\section*{\label{sec:Contribution}Author contributions}
X.Z. conceived the idea and designed the research. X.Z. and X.R. performed the experiments with assistance of X.S. X.Z. designed the device. X.Z. and D.X. fabricated the device. X.Z., X.R., X.S., D.X., and X.W. developed the test equipments. X.Z., H.J., and J.Z. conducted the theory with advice from F.N. and C.-W.Q. X.Z. and H.J. wrote the manuscript with inputs from all authors. The project was jointly supervised by X.Z. and  H.J.

\section*{\label{sec:Competing}Competing interests}
The authors declare no competing interests.

\end{document}